\documentstyle {article}
\title          {SUPERFIELD EFFECTIVE POTENTIAL}
\author{I.L.Buchbinder, S.M.Kuzenko, A.Yu.Petrov and J.V.Yarevskaya\\
        Department of Theoretical Physics\\
        State Pedagogical Institute\\
        Tomsk 634041, Russia}
\date{}
\begin {document}
\maketitle
                                                       TSPI-TH2/94
\\
\\
\\
\\
\\
\\
\\
\\
\\
\\
{\it To be published in the Proceedings of the Conference on
 Heat Kernel Technique and Quantum Gravity, Winnipeg, Canada,
 August 2-6, 1994}

\newpage
       {\huge\bf  SUPERFIELD EFFECTIVE POTENTIAL}\\
         \\
        I.L.Buchbinder, S.M.Kuzenko, A.Yu.Petrov and I.V.Yarevskaya\\
        Department of Theoretical Physics\\
        Tomsk State Pedagogical Institute\\
        Tomsk 634041, Russia\\
                                                       TSPI-TH2/94
\\
\section{Introduction}

This paper is a brief review of recent work carried out by our
group [1-3] on definition and calculation of effective potential
in N=1 supersymmetric field theory.
As is well known, many problems of quantum field theory lead to
necessity of studing of effective action or, in certain cases,
effective potential. It is quite natural to expect that in theories
processing some invariance group there should exist manifestly
covariant methods for determing these objects in an explicitly
symmetric form.
The most adequate formulation for the N=1 supersymmetric field
theories is based on the use of the N=1 superspace and corresponding
superfields (see for example [4-6]). Any physical quantity of inte-
rest arising in such theories may be described in superfield terms.
As to the effective potential in supersymmetric field theories, a
problem of its calculation was studied by many authors [7-14] but
practically all results have been obtained at the component level.
The final form of superfield effective potential was unknown till
nowadays.
An essence of problem from our point of view looks like as follows.
In convential field theory the effective potential is an effective
lagrangian evaluated at constant values of scalar fields, so its
calculated can be implemented with the help of standard methods
\cite{Coleman}. If we try to define the superfield effective
potential as the superfield effective action at constant chiral
and antichiral scalar superfields we will obtain that the effective
action vanishes because of the known properties of the Berezin
integral. Thus we must consider the superfield effective action
for the scalar superfields preserving the arbitrary dependence
on Grassmann coordinates $\theta$ and $\bar{\theta}$. It means,
for calculation of superfield effective potential it is necessary
to use a propagator of the theory in external superfield nontrivially
depending on $\theta$ and $\bar{\theta}$. It is not clear from the very
beginning that such a propagator can be found in explicit form.
As shown in our papers \cite{Buch1,Buch2,Buch3} to calculate the
superfield effective potential we should evaluate the superfield
effective action at special (supersymmetric) conditions
$\partial_a\Phi = \partial_a\bar{\Phi} = 0$
where $\Phi(\bar{\Phi})$ is chiral (antichiral) scalar superfield.
Dependence of the $\Phi$ and $\bar{\Phi}$ on $\theta$ and $\bar{\theta}$
is arbitrary. Thus we have a theory in the external superfields
$\Phi$ and $\bar{\Phi}$ of above type. In thus case the standard
super graph technique which is very convenient for conterterm
computations turns out to be badly adapted for treating the superfield
effective potential. The more efficient approach should be developed.
We have shown that the problem under consideration can be solved
very efficienty in framework of superfield proper time technique.
As a result we have found that the superfield effective potential
is determined by three effects: k\"ahlerian effective potential,
chiral effective potential and auxiliary fields effective potential.
The main role for calculation of all these objects plays the super
field analog of heat kernal which can be solved explicitly for the
arbitrary $\theta$, $\bar{\theta}$ - dependent background superfields.
The question about chiral effective potential is worthly of special
discussion. According to nonrenormalization theorem
\cite{Gates,Buch4,Grisaru2},
all loop corrections to the effective action are expressed in terms of
integrals over whole superspace but not over its chiral subspace.
It was commonly believed as a result that the super symmetric
effective potential does not get chiral-line quantum corrections,
but it is not time, It was shown by West \cite{West1} that in massless
supersymmetric theories can arise contributions to the effective
action of the form
\begin{equation}
\int d^4x d^4 \theta g(\Phi) (-\frac{D^2}{4\Box})f(\Phi)
\end{equation}
which can be written due to the chirality of $\Phi$ in the form
\begin{equation}
\int d^4x d^2 \theta g(\Phi) f(\Phi)
\end{equation}
and presents a contribution to chiral potential. This observation\\
was further confirmed in ref \cite{Buch3,Jack,Dunbar,West2}.
\newpage
\section{The structure of effective action for Wess-Zumino model.}

In this section we are going to discuss the general properties of\\
superfield effective action of the Wess-Zumino (WZ) model. Its\\
action has the form
\begin{equation}
 S[\Phi,\bar{\Phi}]=\int d^8 z\Phi \bar{\Phi} +
                   (\int d^6 z {\cal L}_c (\Phi)+h.c.)
\end{equation}
\begin{equation}
 {\cal L}_c (\Phi) = \frac{m}{2}\Phi^2 +
                     \frac{\lambda}{3!}\Phi^3, \bar{D}_{\dot{\alpha}}\Phi=0
\end{equation}
i.e this is theory of chiral superfield.\\
The effective action of WZ model $\Gamma [\Phi, \bar{\Phi}]$ has the common\\
definition as the Legendre transform of the generating functional
for connected Green function $W[J,\bar{J}]$:
\begin{eqnarray}
\label{Green}
 \exp(\frac{i}{\hbar}W[J,\bar{J}]) & = & \int {\cal D} \Phi {\cal D} \bar{\Phi}
\exp(\frac{i}{\hbar}S[\Phi,\bar{\Phi}]+J\Phi+\bar{J}\bar{\Phi})\nonumber\\
 \Gamma[\Phi,\bar{\Phi}] & = & W[J,\bar{J}]-J\Phi-\bar{J\Phi}
\end{eqnarray}
where $\Phi$ is the vacuum expectation value of $\phi$ in theory with action
$S[\Phi,\bar{\Phi}] + j\phi + \bar{j}\bar{\phi}$.
After extracting of quantum fields $\chi,\bar{\chi}$ we will have
\begin{eqnarray}
 \exp(\frac{i}{\hbar}\bar{\Gamma}[\Phi,\bar{\Phi}]) =
 \int {\cal D} \chi {\cal D} \bar{\chi}
 \exp[iS^{(\psi)}[\chi,\bar{\chi}] \nonumber\\
+ \frac{i}{\sqrt{\hbar}}
 (\frac{\hbar\lambda}{3!} \int d^6z \chi^3 -
  \chi \frac{\delta\bar{\Gamma}}{\delta\Phi} + h.c.)]
\end{eqnarray}
where $ \bar{\Gamma}[\Phi,\bar{\Phi}] =
             \Gamma[\Phi,\bar{\Phi}] - S[\Phi,\bar{\Phi}]$ and
\begin{eqnarray}
 S^{(\psi)}[\chi,\bar{\chi}] & = & \int d^8z\chi\bar{\chi} -
   (\frac{1}{2}\int d^6z\psi\chi^2 + h.c.)\nonumber\\
 - \psi(z) & = & m + \lambda\Phi(z) = {\cal L}_c ''(\Phi),
\bar{D}_{\dot{\alpha}}\Phi=0.
\end{eqnarray}
We will carry out our calculations in the framework of loop expansion
\begin{equation}
\label{Gamma}
 \Gamma[\Phi,\bar{\Phi}] =
\sum_{n=1}^{\infty}\hbar^n\Gamma^{(n)}[\Phi,\bar{\Phi}]
\end{equation}
At the one-loop level, one gets
\begin{equation}
\label{Gamma(1)}
 \Gamma^{(1)}[\Phi,\bar{\Phi}] = -\frac{i}{2}Tr \ln G^{(\psi)}
\end{equation}
At the two-loop level one gets
\begin{equation}
 \Gamma^{(2)}[\Phi,\bar{\Phi}] = -\frac{i}{2}\int{\cal D}\chi{\cal D}\bar{\chi}
               e^{iS^{(\psi)[\chi,\bar{\chi}]}} S^2_{int}[\chi,\bar{\chi}]
\end{equation}
$G^{(\psi)}$ is a matrix superpropagator for theory (7),
\begin{equation}
 G^{(\psi)}(z_1,z_2) =
 \left(
 \begin{array}{ll}
  G_{++}(z_1,z_2) & G_{+-}(z_1,z_2)\\
  G_{-+}(z_1,z_2) & G_{--}(z_1,z_2)
 \end{array}
 \right)
\end{equation}
it satisfies the equation
\begin{equation}
\label{def}
 -\left(
 \begin{array}{cc}
  \psi & \frac{1}{4}\bar{D}^2\\
  \frac{1}{4}D^2 & \bar{\psi}
 \end{array}
 \right)
 \left(
 \begin{array}{ll}
  G_{++} & G_{+-}\\
  G_{-+} & G_{--}
 \end{array}
 \right)
 = \left(
 \begin{array}{ll}
  \delta_+ & 0\\
  0 & \delta_-
 \end{array}
 \right)
\end{equation}
Calculation of $\Gamma^{(1)}$ is a very difficult problem due to very
complicated structure of $G^{(\psi)}$. But using the results of [1-2] we can
obtain the expression for $\Gamma^{(1)}[\Phi,\bar{\Phi}]$ in terms
of the Green function $G^{(\psi)}_v$ satisfying the equation
\begin{eqnarray}
 \Delta G^{(\psi)}_v(z_1,z_2) = -\delta^8 (z_1-z_2)\nonumber\\
 \Delta = \Box + \frac{1}{4}\psi\bar{D}^2 + \frac{1}{4}\bar{\psi}D^2
\end{eqnarray}
We can rewrite $\Gamma^{(1)}[\Phi,\bar{\Phi}] = -\frac{i}{2} tr \ln G^{\psi}_v$
The expression for $G^{\psi}_v$ has the form
\begin{equation}
\label{Green1}
 G^{(\psi)}(z_1,z_2) = \frac{1}{16}
 \left[
 \begin{array}{ll}
  \bar{D}^2_1\bar{D}^2_2 G_{++}(z_1,z_2) & \bar{D}^2_1 D^2_2 G_{+-}(z_1,z_2)\\
  D^2_1\bar{D}^2_2 G_{-+}(z_1,z_2)       & D^2_1 D^2_2 G_{--}(z_1,z_2)
 \end{array}
 \right]
\end{equation}
For calculation of $G^{(\psi)}_v$ we can use proper-time representation
\begin{equation}
 G^{(\psi)}_v = i \int_0^{\infty} ds U^{(\psi)}_v(z_1,z_2;s)
\end{equation}
where the kernel $U^{(\psi)}_v$ is expressed as
\begin{eqnarray}
\label{U}
 U^{(\psi)}_v(z_1,z_2;s) & = & \exp(is\Delta)\delta^8(z_1-z_2)\\
 U_v(s) = U^{(0)}_v(s)   & = & - \frac{i}{(4\pi s)^2}\delta_{12}
                               \exp[\frac{i}{4}\frac{(x-x')^2}{s}] \nonumber
\end{eqnarray}
Then \ref{Gamma(1)} can be rewritten in the form
\begin{equation}
 \Gamma^{(1)}[\Phi,\bar{\Phi}] = -\frac{i}{2} \int_0^{\infty}
                                  \frac{ds}{s} tr U^{(\psi)}_v(s)
\end{equation}
We also can read off the general structure of the effective action
\begin{equation}
 \Gamma[\Phi,\bar{\Phi}] = \int d^8z L(\Phi,D_A\Phi,D_AD_B\Phi,\ldots
 \bar{\Phi},D_A\bar{\Phi},D_AD_B\bar{\Phi}) + (\int d^6z L_c(\Phi) + h.c.)
\end{equation}
L is effective super lagrangian
\begin{eqnarray}
 L & = & K(\Phi,\bar{\Phi}) +
F(D_{\alpha}\Phi,D^2\Phi,\bar{D}_{\dot{\alpha}}\bar{\Phi},
         \bar{D}^2\bar{\Phi},\Phi,\bar{\Phi})+\ldots \nonumber \\
 K & = & \Phi\bar{\Phi} + \sum_{n=1}^\infty \hbar^n K^{(n)}\\
 F & = & \sum_{n=1}^\infty \hbar^n F^{(n)}\nonumber\\
         F\big|_{D_{\alpha}\Phi=\bar{D}_{\dot{\alpha}}=0} \nonumber
\end{eqnarray}
K is the k\"ahlerian effective potential (depending only on fields
but not of their derivatives).
F is the auxiliary fields potential (it is at least of third order
in auxiliary fields of $\Phi$ and $\bar{\Phi}$).
$L_c$ is the chiral effective potential (one-loop chiral contributi
on is equal to zero but higher corrections exist).
The effective potential of WZ model is defined to be the effective
lagrangian evaluated at the superfield conditions
\begin{equation}
\label{superfield}
 \partial_a\Phi = \partial_a\bar{\Phi} = 0
\end{equation}
To calculate K and F it is sufficient to evaluate the effective ac
tion for superfield satisfying eq.(20). As to $L_c$,
we can calculate $\bar{\Psi}$-independent part of $G^{(\psi)}_v$
straightforward, and we have (\cite{Buch1,Buch2})
\begin{equation}
\label{Green2}
 G^{(\psi)}_v(z_1,z_2) = -\frac{1}{\Box} \delta^8(z_1-z_2) +
 \frac{1}{4\Box}(\psi(z_1)\frac{\bar{D}^2}{\Box}\delta^8(z_1-z_2)) +
 O(\bar{\Psi})
\end{equation}
This ansatz may be used for loop calculations of $L_c$.
\newpage
\section{One-loop approximation}

To calculate one-loop effective potential we must calculate the heat
kernel $U^{(\psi)}_v$ for special values of the background superfields:
\begin{equation}
 \partial_a\Phi = \partial_a\bar{\Phi} = 0
\end{equation}
(\ref{U}) leads to
\begin{equation}
 U^{(\psi)}_v(s) = \exp [\frac{i}{4}s (\psi\bar{D}^2 + \bar{\psi}D^2)]
                   U_v(s) = \Omega(\psi|s)U_v(s)
\end{equation}
To calculate $\Omega(\psi|s)$ we will use the Schwinger---De Witt
method: really $\Omega(\psi|s)$ satisfies the equation
\begin{equation}
 i\frac{\partial\Omega}{\partial{s}} + \frac{1}{4}
                \Omega(\psi\bar{D}^2 + \bar{\psi}D^2) = 0
\end{equation}
with initial condition
$$ \Omega(0|s) = 1$$
We will represent $\Omega(\psi|s)$ in the form
\begin{eqnarray}
 \Omega(\psi|s) = 1 + \frac{1}{16}A(s)D^2\bar {D}^2 +
 \frac{1}{16}\bar{A}(s)\bar{D}^2D^2 +
 \frac{1}{8}B^{\alpha}(s)D_{\alpha}\bar{D}^2 +
 \frac{1}{8}\bar{B}_{\dot{\alpha}}(s)\bar{D}^{\dot{\alpha}}D^2 +\nonumber\\
 \frac{1}{4}C(s)D^2 + \frac{1}{4}\bar{C}(s)\bar{D}^2
\end{eqnarray}
and we will obtain equations for coefficients $A,\tilde{A},B,\tilde{B},C,\tilde{C}C$.
Since we are interested to calculate the trace of $U^{(\psi)}_v$, we must to
calculate essentially $A$ and $\tilde{A}$, because
\begin{equation}
 tr U^{(\psi)}_v(s) = \int d^8z \int d^4x' \delta^4(x-x')[A(s)+\tilde{A}(s)]U(x,x';s)
\end{equation}
due to famous properties of trace and covariant derivatives.
To solve the system of equations for coefficients $A,\tilde{A},B,\tilde{B},C,\tilde{C}$
is a very complicated task, but for calculating k\"ahlerian potenti
al (when $\psi$ and $\bar{\psi}$ are constant) the system is simplified and
 kernel has the form [1-2]
\begin{eqnarray}
 U^{\psi=const}_v(s) = (1 + \frac{1}{16\Box}
 [\cosh(is\sqrt{\psi\bar{\psi}\Box}-1]\times\nonumber\\
 \{D^2,\bar{D}^2\} +
 \frac{1}{4\sqrt{\psi\bar{\psi}\Box}}
 \sinh(is\sqrt{\psi\bar{\psi}\Box})(\psi\bar{D}^2+\bar{\psi}D^2))U_v(s)
\end{eqnarray}
To calculate regularized k\"ahlerian potential we must put term pro
portional to $\{D^2,\bar{D}^2\}$ and will obtain
\begin{equation}
 K^{(1)}_{reg} = - i \int_{L^2}^{\infty} \frac{ds}{s} \int d^4x' \delta^4(x-x')
                \frac{1}{\Box} [\cosh(is\sqrt{\psi\bar{\psi}\Box}-1] U(x,x';s)
\end{equation}
(We must use regularization by means of the proper-time cutoff be
cause we have singularity at lower limit). Using famous properties
of delta-function and kernel $U(x,x';s)$ we can obtain
\begin{equation}
 K^{(1)}_{reg} = - \frac{1}{32\pi^2}\psi\bar{\psi}
               (\ln\frac{\psi\bar{\psi}}{\mu^2} - \xi + \ln \mu^2L^2)
\end{equation}
choosing the one-loop wave-function renormalization
\begin{equation}
 Z^{(1)} = \frac{\lambda^2}{32\pi^2} \ln \mu^2L^2
\end{equation}
cancels divergences, and we stay with one-loop effective k\"ahlerian
potential
\begin{equation}
\label{K1}
 K^{(1)}(\Phi,\bar{\Phi}) = -
\frac{1}{32\pi^2}(m+\lambda\Phi)(m+\lambda\bar{\Phi})
        [\ln\frac{(m+\lambda\Phi)(m+\lambda\bar{\Phi})}{\mu^2} - \xi]
\end{equation}
As to auxiliary fields' potential, it does not contain divergences,
but its explicit calculation is very complicated due to highly cum
bersome structure of the kernel $U^{(\psi)}_v(s)$, and we can only
to develop procedure for perturbative determining $F^{(1)}$ (we ha
ve not yet succeeded in bringing result for auxiliary fields' pot
ential in final form). For example, in lower (fourth) order we get
\begin{equation}
 F^{(1)} = \frac{\lambda^4\zeta}{(4\pi)^2} (D^{\alpha}\Phi)(D_{\alpha}\Phi)
 (\bar{D}_{\dot{\alpha}}\bar{\Phi})(\bar{D}^{\dot{\alpha}}\bar{\Phi})
 |{\cal L}''_c(\Phi)|^{-4} + O^8(D,\bar{D})
\end{equation}
where $\zeta$ is a finite constant, defined in \cite{Buch1}.
\newpage
\section{Two-loop approximation}

Calculation of two-loop chiral and k\"ahlerian effective potential
is based on properties of Green function (\ref{def}),(\ref{Green1}),
(\ref{Green2}) and presents no difficulties. For calculating two-loop
chiral effective potential we must use (\ref{Green1}) and (\ref{Green2})
and will get $\bar{\psi}$-independent part of $G^{(\psi)}$:
\begin{equation}
\label{Matrix}
 G^{(\psi)}(z_1,z_2) = \frac{1}{16}
 \left[
  \begin{array}{cc}
  0 & -\frac{\bar{D}^2_1D^2_2}{\Box}\delta^8(z_1-z_2)\\
  -\frac{D^2_1\bar{D}^2_2}{\Box}\delta^8(z_1-z_2) &
  -\frac{D^2_1D^2_2}{\Box}\psi(z_1)\frac{\bar{D}^2_1}{4\Box_1}\delta^8(z_1-z_2)
  \end{array}
 \right]
\end{equation}
It follows from (\ref{Green}), (\ref{Gamma}), (\ref{Matrix}) that the two-loop
contribution to $\Gamma[\Phi,\bar{\Phi}]$ depending only on $\psi$ looks like
\begin{eqnarray}
 \Gamma^{(2)}    & = & \frac{\lambda^2}{12} \int d^6\bar{z}_1
                       d^6\bar{z}_2 [G_{--}(z_1,z_2)]^3 \nonumber \\
 G_{--}(z_1,z_2) & = & \frac{D^2_1D^2_2}{16\Box_1}(\psi(z_1)
                       \frac{\bar{D}^2_1}{4\Box_1}\delta^8(z_1-z_2))
\end{eqnarray}
After quite obvious transformation we will get for massless case
\begin{eqnarray}
 \Gamma^{(2)} = -\frac{\lambda^5}{12} \int \prod_{i=1}^5 d^8z_i
 \Phi(z_3)\Phi(z_4)\Phi(z_5)\nonumber\\
\frac{1}{\Box_1}\delta^8(z_1-z_3)
 \frac{D^2_2\bar{D}^2_3}{16\Box_2}\delta^8(z_3-z_2) \times
\frac{1}{\Box_2}\delta^8(z_2-z_4)\frac{D^2_1\bar{D}^2_4}{16\Box_1}\delta^8(z_1-z_4)\nonumber\\
\frac{D^2_1\bar{D}^2_5}{16\Box_1}\delta^8(z_1-z_5)\frac{D^2_2}{4\Box_2}\delta^8(z_2-z_5)
\end{eqnarray}
After D-algebra transformations and converting to momentum representation
 we arrive at
\begin{eqnarray}
 \Gamma^{(2)} = -\frac{\lambda^5}{12} \int d^4xd^4\theta
\int\frac{d^4k_1d^4k_2d^4p_1d^4p_2}{(2\pi)^{16}}
\int d^4y_1d^4y_2 e^{ip_1(x-y_1)+ip_2(x-y_2)} \Phi(x,\theta) \times\nonumber\\
\relax
[k^2_1p^2_1(-\frac{D^2}{4\Box}\Phi(y_1,\theta))\Phi(y_2,\theta) +
(1\leftrightarrow 2) + \frac{1}{2}
(k_1,k_2) D^{\alpha} \Phi(y_1,\theta) D_{\alpha} \Phi(y_2,\theta)]\nonumber\\
\Omega^{-1}(k,p)
\end{eqnarray}
where
\begin{equation}
 \Omega(k,p)=k^2_1k^2_2(k_1+k_2)^2(k_1-p_1)^2(k_2-p_2)^2(k_1+k_2-p_1-p_2)^2
\end{equation}
Then use the rule $\int d^4\theta=\int d^2\theta(-\frac{1}{4}D^2)$ we will get
\begin{eqnarray}
 \Gamma^{(2)}=\frac{\lambda^5}{12}\int d^4x d^2\theta d^4y_1
d^4y_2e^{ip_1(x-y_1)+ip_2(x-y_2)}\times
 \Phi(x,\theta)\Phi((y_1,\theta)\Phi(y_2,\theta)\times\nonumber\\
J(p_1,p_2)
\end{eqnarray}
where
\begin{equation}
 J(p_1,p_2)=\int \frac{d^4k_1 d^4k_2}{(2\pi)^8}
 \frac{k^2_1 p^2_1+k^2_2 p^2_2-2k_1k_2p_1p_2}{\Omega(k,p)}
\end{equation}
To calculate effective potential we must put $\Phi$ to be slowly varying
in spacetime, it implies
\begin{equation}
\label{res}
 \Phi(x,\theta)\Phi(y_1,\theta)\Phi(y_2,\theta) \simeq \Phi^3(x,\theta)
\end{equation}
Then \ref{res} takes the form
\begin{equation}
 \Gamma^{(2)}=\frac{\lambda^5}{12}J(p_1\rightarrow 0,p_2\rightarrow 0)
\int d^6z\Phi^3(z)
\end{equation}
It is famous that
\begin{equation}
J(p,0)=\frac{6}{(4\pi)^4}\zeta(3)
\end{equation}
and the two-loop chiral effective superpotential reads as [3]
\begin{equation}
 L_c(\Phi) = {\cal L}_c + \hbar^2{\cal L}^{(2)}_c = (\frac{\lambda}{3!} +
\frac{\lambda^5\zeta(3)\hbar^2}{2(4\pi)^4})\Phi^3
\end{equation}
As for two-loop k\"ahlerian potential, we can calculate $G^{(\psi)}$ for con-
stant values and obtain
\begin{equation}
 G^{(\psi)} = -\frac{1}{\Box-M^2}
 \left(
 \begin{array}{cc}
  \bar{\psi} & -\frac{1}{4}\bar{D}^2\\
  -\frac{1}{4}D^2 & \psi
 \end{array}
 \right)
 \left(
 \begin{array}{ll}
  \delta_+ & 0\\
  0 & \delta_-
 \end{array}
 \right)
 + \ldots
\end{equation}
where $M^2 = \psi\bar{\psi}$, dots means all terms which give no contribution
to
$K^{(2)}$.
   From \ref{Gamma} we can conclude that $K^{(2)}$ is expressed by
Green function as
\begin{equation}
 K^{(2)} = \frac{\lambda^2}{6} \int d^6z_1d^6z_2(G_{+-}(z_1,z_2))^3
\end{equation}
In momentum representation it leads to
\begin{eqnarray}
 K^{(2)} = -\frac{\lambda^2}{6} \int d^4xd^4\theta_1d^4\theta_2
            \frac{d^4kd^4l}{(2\pi)^8}\frac{1}{(k^2+M^2)(l^2+M^2)[(k+l)^2+M^2]}
            \nonumber\\
            \frac{D^2_1}{4}\delta_{12}\frac{D^2_1\bar{D}^2_2}{16}
            \delta_{12}\frac{\bar{D}^2_2}{4}\delta_{12}
\end{eqnarray}
After D-algebra transformations, integration by momenta and regula-\\
rization we have
\begin{eqnarray}
 K^{(2)}_{REG} = \frac{\lambda^2}{6(4\pi)^4} \int d^4\theta
[\frac{6M^2}{\epsilon^2}+\frac{3M^2}{\epsilon}(3-2\gamma-2\ln\frac{M^2}{\mu^2})+\nonumber\\
 3M^2\ln^2\frac{M^2}{\mu^2}-3M^2(3-2\gamma)\ln\frac{M^2}{\mu^2}+9M^2(1-\gamma)]
\end{eqnarray}
After subtraction of one-loop and two-loop counterterms we arrive to
\begin{eqnarray}
 K^{(2)}_{renorm} = -\frac{\lambda^2}{(4\pi)^4}
 \int d^4\theta (m+\lambda\Phi)(m+\lambda\bar{\Phi})
[-\frac{1}{4}\ln^2\frac{(m+\lambda\Phi)(m+\lambda\bar{\Phi})}{\mu^2}\nonumber\\
 +\frac{3-\gamma}{2}\ln\frac{(m+\lambda\Phi)(m+\lambda\bar{\Phi})}{\mu^2} +
  \frac{3}{2}(\gamma-1) + \frac{1}{4}(\gamma^2+\zeta(2)) - b]
\end{eqnarray}
where b is finite part of two-loop counterterm. It will be determined
by renormalization condition (for massive case)
\begin{equation}
 \frac{\partial^2}{\partial\Phi\partial\bar{\Phi}}(K^{(0)} + K^{(1)} + K^{(2)}
\big|_{\Phi=\bar{\Phi}=0}=1
\end{equation}
where $K^{(0)}=\Phi\bar{\Phi}$ and $K^{(1)}$ is given in (29).
After calculations we get
\begin{eqnarray}
 K^{(2)} = \frac{\lambda^2}{(4\pi)^4}
 \int d^4\theta (m+\lambda\Phi)(m+\lambda\bar{\Phi})
 [\frac{1}{4}\ln^2\frac{(m+\lambda\Phi)(m+\lambda\bar{\Phi})}{m^2} -\nonumber\\
  \frac{5}{2}\ln\frac{(m+\lambda\Phi)(m+\lambda\bar{\Phi})}{m^2} + \frac{9}{2}]
\end{eqnarray}
and using this choice for $\mu$ we have
\begin{equation}
 K^{(1)} = -\frac{1}{2(4\pi)^2}
 \int d^4\theta (m+\lambda\Phi)(m+\lambda\bar{\Phi})
 [\ln\frac{(m+\lambda\Phi)(m+\lambda\bar{\Phi})}{m^2} -2]
\end{equation}

\end{document}